\documentclass[graybox]{svmult}

\usepackage{mathptmx}       
\usepackage{helvet}         
\usepackage{courier}        
\usepackage{type1cm}        
\usepackage{amssymb}
\usepackage{amsmath} 
\usepackage{makeidx}         
\usepackage{graphicx}        
\usepackage{multicol}        
\usepackage[bottom]{footmisc}

\makeindex             
                       
\makeatletter
\renewcommand{\fnum@figure}{\textbf{\figurename~\thefigure}}
\renewcommand{\fnum@table}{\textbf{\tablename~\thetable}}

\newcommand{\ba}{\begin{array}{llll}}
\newcommand{\ea}{\end{array}}

\newcommand{\be}{\begin{equation}}
\newcommand{\ee}{\end{equation}}

\newcommand{\bi}{\begin{itemize}}
\newcommand{\ei}{\end{itemize}}

\newcommand{\bc}{\begin{center}}
\newcommand{\ec}{\end{center}}

\newcommand{\bfig}{\begin{figure}[!ht]}
\newcommand{\efig}{\end{figure}}

\newcommand{\ben}{\begin{enumerate}}
\newcommand{\een}{\end{enumerate}}

\newcommand{\bmat}{\left[\begin{matrix}}
\newcommand{\emat}{\end{matrix}\right]}

\begin{document}

\title*{Collision-free speed model for pedestrian dynamics}
\author{Antoine Tordeux \and Mohcine Chraibi \and Armin Seyfried}
\institute{Antoine Tordeux \at J\"ulich Supercomputing Centre, Forschungszentrum J\"ulich GmbH, Germany and Computer Simulation for Fire Safety and Pedestrian Traffic, Bergische Universit\"at Wuppertal, Germany, \email{a.tordeux@fz-juelich.de}
\and Mohcine Chraibi \at J\"ulich Supercomputing Centre, Forschungszentrum J\"ulich GmbH, Germany \email{m.chraibi@fz-juelich.de}
\and Armin Seyfried \at J\"ulich Supercomputing Centre, Forschungszentrum J\"ulich GmbH, Germany and Computer Simulation for Fire Safety and Pedestrian Traffic, Bergische Universit\"at Wuppertal, Germany, \email{a.seyfried@fz-juelich.de}}
%
%
\maketitle

\abstract{We propose in this paper a minimal speed-based pedestrian model for which 
particle dynamics are intrinsically collision-free. 
The speed model is an optimal velocity function depending on  
the agent length (i.e.\ particle diameter), maximum speed and time gap parameters. 
The direction model is a weighted sum of exponential repulsion from the neighbors,  
calibrated by the repulsion rate and  distance. 
The model's main features like the reproduction of empirical phenomena are analysed by simulation. 
We point out that phenomena of self-organisation observable in force-based models and field studies 
can be reproduced by the collision-free model with low computational effort.}

\section{Introduction} \label{intro}
Modelling of pedestrian dynamics have been strongly developed since the 1990's 
\cite{Chowdhury2000,Schadschneider2010a,Degond2013}. 
Microscopic models describe the movement of individuals in two-dimensional representation of space. 
They are used for theoretical purposes \cite{Helbing2006,Helbing2000z}, as well as for applications e.g.~design and conception of escape routes in buildings~\cite{Schneider2002,TraffGo2005} 
or optimal organization of mass events or public transport facilities (VISWalk~\cite{ptv}, Legion~\cite{legion}, $\ldots$). 
In the microscopic class of models, pedestrians are represented as autonomous entities (Lagrangian representation) with local interactions.   
Complex collective phenomena of self-organisation emerge from the interactions. 
Examples are the lane formation, clogging at bottlenecks, zipper effect or intermittent flow at bottlenecks, 
stop-and-go waves, herding, strip formation or circular flows (see \cite{Chowdhury2000,Helbing2001} and references therein). 
Even simple microscopic models can yield in rich dynamics \cite{Helbing1995,Chraibi2010a}.
Yet, the relations between the microscopic model parameters and 
the emergence of phenomena of self-organisation  are not straightforward.
In most of the cases, they have to be analysed by simulation. 

Microscopic pedestrian models could be defined in continuous or discrete time, space and state variables 
(see \cite[Chapter~5]{Schadschneider2010a}). 
One of the most investigated class is the class of \emph{force-based} (or acceleration) models \cite{Helbing1995,Chraibi2010a,Chraibi2011}. 
They use an analogy between pedestrian movement and Newtonian dynamics. 
Force-based approaches allow to describe a large variety of pedestrian dynamics \cite{Helbing1995,Chraibi2010a}. 
Yet, this model class describes particles with inertia and does not exclude particle collision and overlapping. 
This is especially problematic at high densities \cite{Chraibi2011}. 
Moreover, the force-based approach may lead to numerical difficulties resulting 
in small time steps and high computational complexity, or use of mollifies \cite{Koester2013}. 

Pedestrian behaviors result from repulsive and attractive forces with the acceleration models.  
They are based on the visual perception of distances or obstacle speeds resulting in instantaneous changing of the speed or the direction 
within the speed models.
Also, this model class is generally called \emph{vision-based}. 
One example is the \emph{synthetic-vision-based steering approach} 
that notably allows to describe complex collective structures avoiding gridlocks \cite{Ondrej2010}. 
Also the \emph{velocity obstacle models} or \emph{reciprocal velocity obstacle model} borrowed from robotics exist \cite{Fiorini1998,Berg2008}. 
These models are defined in discrete time and are driven by collision avoidance. 
They are by construction collision-free if the time step is smaller than a horizon time of anticipation. 
In the evacuation model by Venel, the pedestrians move as fast as possible to the desired destination with no overlapping \cite{Maury2007}. 
There exits some variants of the model with different interaction strategies \cite{Venel2010}. 
Note that there exists also rule based multi-agent models aiming to describe pedestrian psychology 
(see for instance \cite{Pelechano2005a,Guo2010}) or 
mixed models, see for instance the \emph{gradient navigation model} 
where the direction model is defined at first order while the speed is of second order \cite{Dietrich2014}.
In most of cases, these models need a large number of parameters with inherent calibration difficulties and, 
as for force-based models, high computational efforts. 

In this paper, we aim to develop a minimal model for which the dynamics are by construction collision-free (i.e. overlapping-free). 
The model belongs to Maury and Venel mathematical framework \cite{Maury2007}. 
We show by simulation that it allows to describe some expected phenomena of self-organisation 
observed in field studies or in simulations with forced based models.
The model is defined in section~\ref{defmod} while 
the simulation results are presented in section~\ref{sim}. 
Conclusion and working perspective are given in section~\ref{ccl}. 

\section{Collision-free speed-based pedestrian model} \label{defmod}

A continuous speed model is a derivative equation for the velocity. Typical examples are
\be
\dot{\mathbf x}_i=\mathbf v(\mathbf x_i,\mathbf x_j,\ldots)\qquad\mbox{or}\qquad \dot{\mathbf x}_i=V(\mathbf x_i,\mathbf x_j,\ldots)\times\mathbf e_i(\mathbf x_i,\mathbf x_j,\ldots),
\label{modg}\ee
with $\mathbf x_i$ the pedestrian position and $\dot{\mathbf x}_i$ the velocity of pedestrian $i$ (see figure~\ref{schem}). 
The velocity in regulated in one function for the first equality 
while the speed $V$ and the direction $\mathbf e_i$ (unit vector) are regulated separately in the second approach. 

\bfig\bc
\includegraphics{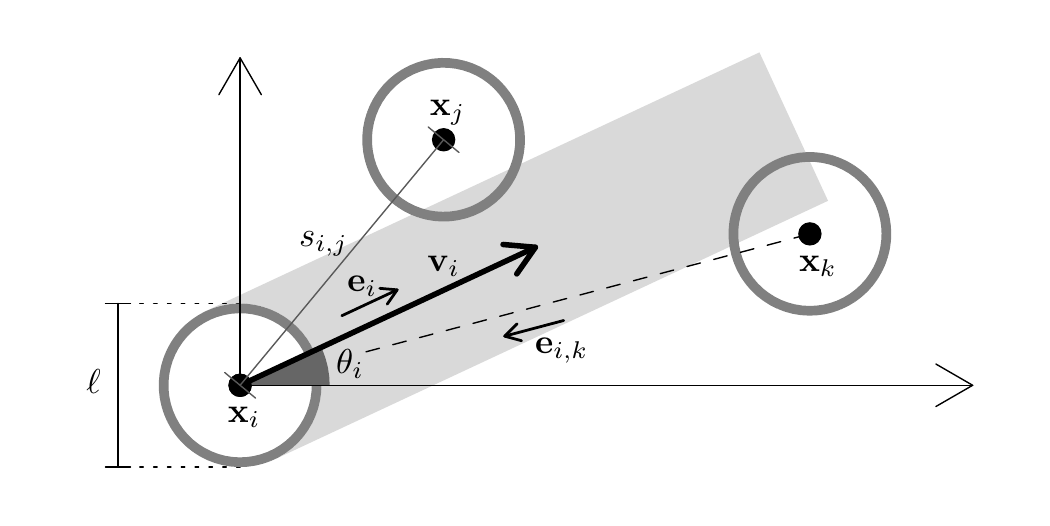}\vspace{-3mm}
\caption{Notations used. $\mathbf x_i$,  $\mathbf v_i$ and $\theta_i$ are the position, velocity and direction of the pedestrian $i\,$; 
$\ell$ is the pedestrian size$\,$; $\mathbf e_{i,j}$ is the unit vector from $\mathbf x_j$ to $\mathbf x_i\,$; 
$\mathbf e_i=(\cos\theta_i,\sin\theta_i)\,$; $s_{i,j}=||\mathbf x_i-\mathbf x_j||$.}
\label{schem}
\ec\efig

\subsection{Definition of the model}

The speed model is the optimal speed (OV) function depending on the minimal spacing in front. 
The approach is borrowed from road traffic model \cite{Bando1995}. 
The OV approach has been already developed with a force-based model \cite{Nakayama2005}. 
Here we use the OV function at the first order with the minimal spacing in front. 

For a given pedestrian $i$, the set of the pedestrians in front is defined by 
\be
J_i=\big\{\,j,\ \mathbf e_i\cdot\mathbf e_{i,j}\le0\ \mbox{ and }\ |\mathbf e_i^\perp\cdot\mathbf e_{i,j}|\le\ell/s_{i,j}\big\}.
\ee
The pedestrians in front are the pedestrians overlapping the grey area in figure~\ref{schem}.
The minimum distance in front $s_i$ is 
\be
s_i=\min_{j\in J_i}s_{i,j}.
\ee
The model is
\be
\dot{\mathbf{x}}_i=V\big(s_i(\mathbf{x}_i,\mathbf{x}_j,\ldots)\big)\times\mathbf e_i(\mathbf{x}_i,\mathbf{x}_j,\ldots),
\label{mod}
\ee
with $V(\cdot)$ the OV function and $\mathbf e_i(\mathbf{x}_i,\mathbf{x}_j,\ldots)$ the direction model to define. 
As shown below, such model is by construction collision-free if
\be
V(s)\ge0\quad\mbox{for all $s$}\qquad\mbox{and}\qquad V(s)=0\quad\mbox{for all $s\le\ell$}.
\label{AV} 
\ee
In the following, the OV function is the piecewise linear $V(s)=\min\{v_0,\max\{0,(s-\ell)/T\}\}$, 
with $v_0$ the desired speed and $T$ the time gap in following situations ($\ell$ is the pedestrian diameter, see figure~\ref{schem}). 
This OV function satisfies the collision-free assumption (\ref{AV}).
The direction model is a simplified version of the additive form of the \emph{gradient navigation model} \cite{Dietrich2014}. 
It is based on a repulsion function depending on the distances $(s_{i,j})$ with the neighbours
\be\textstyle
\mathbf e_i(\mathbf{x}_i,\mathbf{x}_j,\ldots)=\frac{1}{N}\left(\mathbf e_0+\sum_j R(s_{i,j})\,\mathbf e_{i,j}\right),
\label{dmod}
\ee
with $\mathbf e_0$ the desired direction given by a strategic model, 
$N$ a normalization constant such that $\|\mathbf e_i\|=1$ and 
$R(s)=a\,\exp\big((\ell-s)/D\big)$ the repulsion function, calibrated by the
coefficient $a>0$ and distance $D>0$. 
The parameter values used in the simulation are presented in figure~\ref{paramod}.

\bfig\bc\vspace{-2mm}
\includegraphics{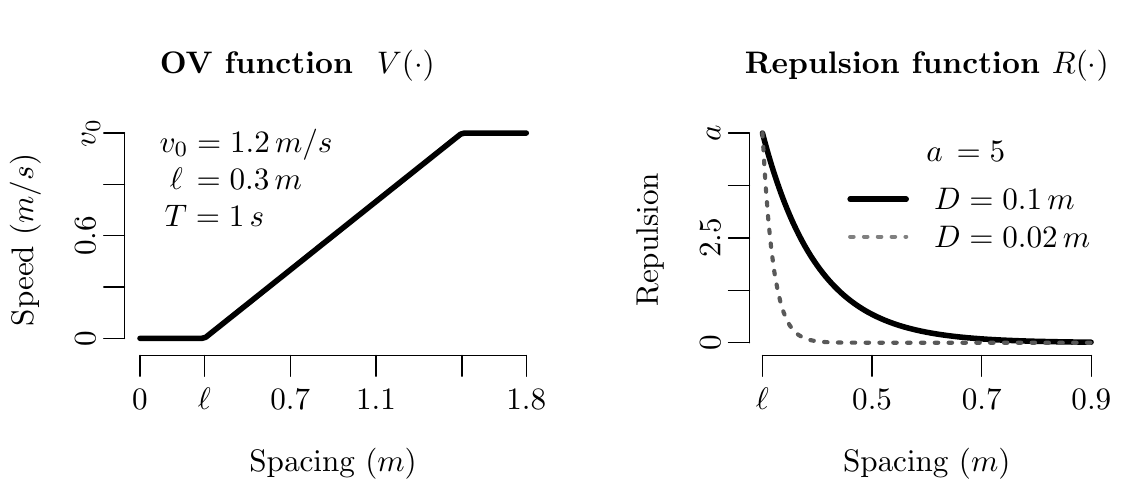}
\caption{Functions and associated parameters for the model: The OV function (3 parameters, left panel), 
and the repulsion function (2 parameters, right panel).}
\label{paramod}\vspace{-1mm}
\ec\efig

\subsection{Collision-free property} 

Oppositely to the force-based models, the presence of collision and overlapping can be controlled by construction with the speed-based models 
(non-overlapping constraint). 
If pedestrians are considered as discs with diameter $\ell$, the set of collision-free configurations is for a given pedestrian $i$
\be
Q_i=\left\{\mathbf x_i\in\mathbb R^{2},\ s_{i,j}\ge\ell\ \ \forall j\right\}.
\ee
The set of collision-free velocities
\be
C_{\mathbf x_i}=\left\{\mathbf v\in\mathbb R^{4},\ s_{i,j}=\ell\ \Rightarrow\  
\mathbf e_{i,j}\cdot \mathbf v_i\ge 0\ \mbox{ and }\ \mathbf e_{j,i}\cdot \mathbf v_j\ge0\right\}
\ee
is such that the speeds are nil or in opposite direction for a pedestrian in contact with an other 
(see \cite{Maury2007} for more general conditions). 
Therefore, if initially $\mathbf x_i(0)\in Q_i$, then $\mathbf x_i$ remains in $Q_i$ for any dynamics in $C_{\mathbf x_i}$. 
In these conditions $Q_i$ is an invariant set for $\mathbf x_i$, i.e. the dynamics are collision-free (see also \cite{Monneau2013}). 
It is easy to see that the model (\ref{mod}) belongs to this class if assumption (\ref{AV}) is satisfied.  
Consider $s_{i,j}=\ell$ then either $\mathbf e_i\cdot\mathbf e_{i,j}\le0$ and 
then $j\in J_i$, i.e. $s_i\le s_{i,j}=\ell$ and $V(s_i)=0$, 
or neither $\mathbf e_i\cdot\mathbf e_{i,j}\ge0$ and then $V(s_i)\ge0$ since $V(\cdot)\ge0$.  
Therefore $\mathbf v_i\cdot\mathbf e_{i,j}=V(s_i)\times\mathbf e_i\cdot\mathbf e_{i,j}\ge0$ 
and the velocity belongs to $C_{\mathbf x_i}$. 
The arguments are valid for any direction model $\mathbf e_i$. 

\section{Model features} \label{sim}

We describe in this section by simulation some characteristics of the model with uni- and bi-directional flows. 
The parameter settings are given in figure~\ref{paramod}. 
The simulations are done on rectangular systems with length $L=9$~m and width $W=3$~m 
from 
random initial configurations and by using explicit Euler numerical scheme with time step $dt=0.01$~s.

\subsection{Counter flows and the lane formation}

We observed with the model the formation of lanes by direction for counter flows (figure~\ref{lane}, left panels). 
Such phenomena frequently occurs in real data (see for instance \cite{ZhangJ2012a}). 
The system needs an organization time for that the lanes emerge (figure~\ref{lane}, 
top right panel), where the mean flow to the desired direction for counter flows is compared to uni-directional ones). 
The formation of lanes is observed with the model for some density levels up to $\rho=6$~ped$/$m$^2$ 
(figure~\ref{lane}, bottom right panel). 
As expected, the density threshold value for that the lanes appear depends on the pedestrian size $\ell$ (here $\ell=0.3$~m).  
Note that the lane formation phenomenon disappears when a noise is introduced in the model 
(freezing by heating phenomenon, see \cite{Helbing2000z} and in figure~\ref{lane}, 
thin dotted line in bottom right panel where a Brownian noise with  
standard deviation $\sigma=0.1$ m$/$s is added to the model -- the lane formation breaks as soon as $\rho\ge2$~ped$/$m$^2$).  

\bfig\bc%
\includegraphics{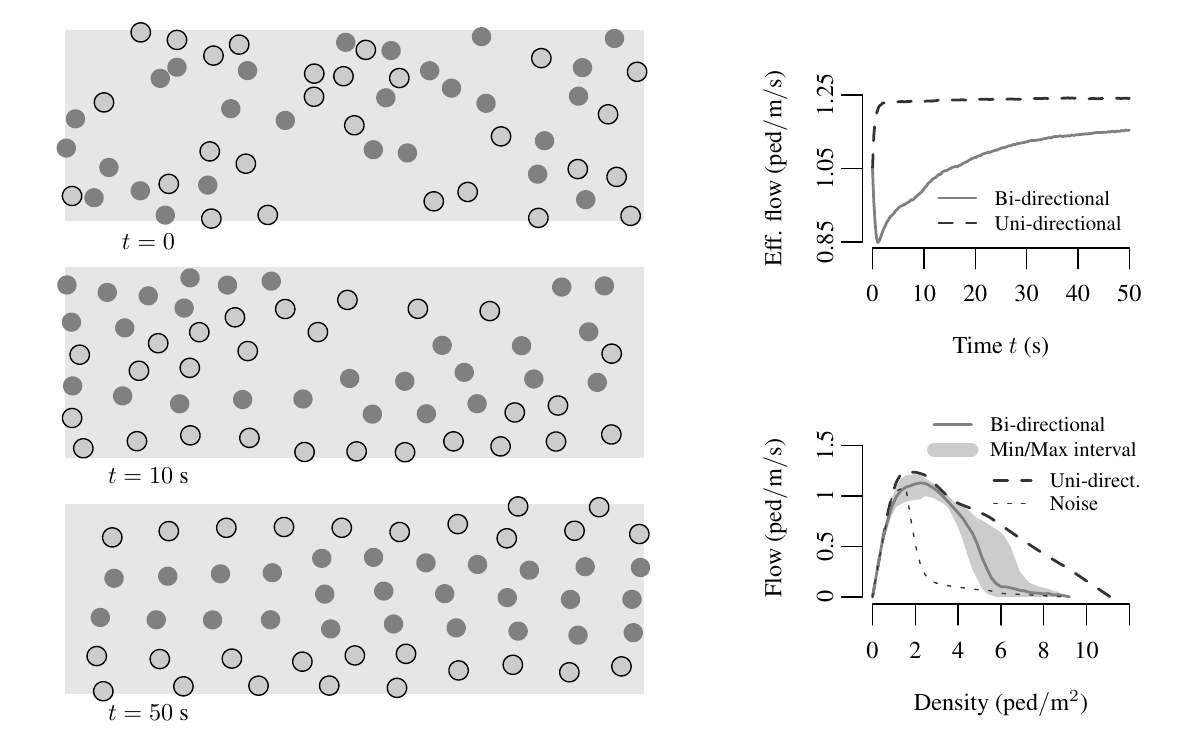}\vspace{-2mm}
\caption{Counter flows. Left panels, snapshots of the system at time $t=0$, $10$ and $20$~s from random initial conditions 
($\rho=2$~ped$/$m$^2$). Right panels, the mean flow sequence to the desired direction 
and the fundamental diagram.}
\label{lane}
\ec\efig

\subsection{Intermittent bottleneck flows}

Oscillating phenomena for counter flows in bottlenecks are observed with both real data and models (\cite{Helbing1995,Helbing2001,Corradi2012}). 
Such phenomena are related as intermittent bottleneck flows in the literature \cite{Helbing2006}. 
We observe that the speed-based model is able to reproduce them (see figure~\ref{int}, left and top right panels). 
The phenomenon occurs even at relatively high density levels (see figure~\ref{int}, bottom right panel). 
Yet it induces frictions and the flow volumes obtained for counter flows are less than the ones of uni-direction. 
As expected, the frictions tend to increase as the density increases. 
Some simulation results not presented here show that the intermittent flow phenomenon subsists for high density levels when 
$D$ is sufficiently high and that the frequency of the flows oscillations tend to increase as the density increases.  

\bfig\bc%
\includegraphics{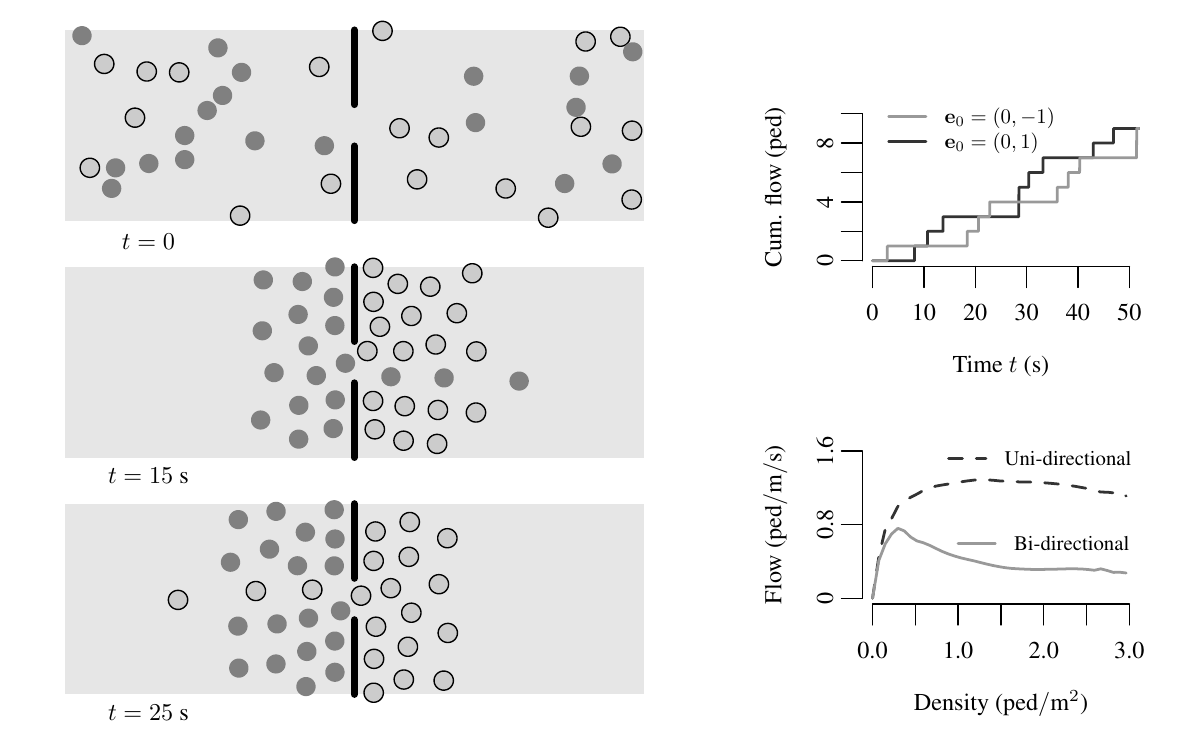}\vspace{-2mm}
\caption{Counter flows with bottleneck. Left panels, snapshots of the system at time $t=0$, $10$ and $20$~s from random initial conditions 
($\rho=1.4$~ped$/$m$^2$ and $\omega=0.6$~m). Right panels, the corresponding flow sequences by direction and the fundamental diagram.}
\label{int}
\ec\efig

\section{Conclusion and working perspective} \label{ccl}

A new speed-based model is proposed for pedestrian dynamics in two dimensions. 
Oppositely to classical force-based approaches, the model is intrinsically collision-free and no 
overlapping phenomena occur, for any density level. 
The model has five parameters. 
Three of them concern the optimal speed function. 
They are the pedestrian length, desired speed and time gap with the predecessor. 
The two others calibrate the direction model. They are the repulsion rate and repulsion distance. 

The model main properties are described by simulation. 
A large range of dynamics observed in real data and force-based models are reproduced. 
For instance, linear increase of flow with the bottleneck width, lane formation for counter flows (with the freezing by heating effect) 
or intermittent flows, are obtained with identical setting of the parameters. 
However, other well-known characteristic such that stop-and-go phenomena can not be described. 
Further mechanisms (and parameters) remain to be introduced to the model.


\begin{thebibliography}{10}
\providecommand{\url}[1]{{#1}}
\providecommand{\urlprefix}{URL }
\expandafter\ifx\csname urlstyle\endcsname\relax
  \providecommand{\doi}[1]{DOI~\discretionary{}{}{}#1}\else
  \providecommand{\doi}{DOI~\discretionary{}{}{}\begingroup
  \urlstyle{rm}\Url}\fi

\bibitem{Bando1995}
Bando, M., Hasebe, K., Nakayama, A., Shibata, A., Sugiyama, Y.: Dynamical model
  of traffic congestion and numerical simulation.
\newblock Phys. Rev. E \textbf{51}(2), 1035--1042 (1995)

\bibitem{Berg2008}
van~den Berg, J., Lin, M., Manocha, D.: Reciprocal velocity obstacles for
  real-time multi-agent navigation.
\newblock In: {Robotics and Automation, 2008. ICRA 2008. IEEE International
  Conference on}. 2008 IEEE International Conference on Robotics and Automation
  Pasadena, CA, USA, May 19-23, 2008 (2008)

\bibitem{legion}
Berrou, J., Beecham, J., Quaglia, P., Kagarlis, M., Gerodimos, A.: Calibration
  and validation of the legion simulation model using empirical data.
\newblock In: N.~Waldau, P.~Gattermann, H.~Knoflacher, M.~Schreckenberg (eds.)
  Pedestrian and Evacuation Dynamics 2005, pp. 167--181. Springer Berlin
  Heidelberg (2007).

\bibitem{Chowdhury2000}
Chowdhury, D., Santen, L., Schadschneider, A.: Statistical physics of vehicular
  traffic and some related systems.
\newblock Phys. Rep. \textbf{329}(4--6), 199--329 (2000).

\bibitem{Chraibi2011}
Chraibi, M., Kemloh, U., Seyfried, A., Schadschneider, A.: Force-based models
  of pedestrian dynamics.
\newblock Netw. Heterog. Media \textbf{6}(3), 425--442 (2011)

\bibitem{Chraibi2010a}
Chraibi, M., Seyfried, A., Schadschneider, A.: Generalized centrifugal force
  model for pedestrian dynamics.
\newblock Phys. Rev. E \textbf{82}, 046,111 (2010).

\bibitem{Corradi2012}
Corradi, O., Hjorth, P., Starke, J.: Equation-free detection and continuation
  of a hopf bifurcation point in a particle model of pedestrian flow.
\newblock SIAM J. Appl. Dyn. Syst. \textbf{11}(3), 1007--1032 (2012)

\bibitem{Degond2013}
Degond, P., Appert-Rolland, C., Moussaid, M., Pett\'e, J., Theraulaz, G.: A
  hierarchy of heuristic-based models of crowd dynamics.
\newblock J. Stat. Phys. \textbf{152}(6), 1033--1068 (2013)

\bibitem{Dietrich2014}
Dietrich, F., K\"oster, G.: Gradient navigation model for pedestrian dynamics.
\newblock Phys. Rev. E \textbf{89}, 062,801 (2014)

\bibitem{Fiorini1998}
Fiorini, P., Shiller, Z.: Motion planning in dynamic environments using
  velocity obstacles.
\newblock Int. J. Robot. Res. \textbf{17}(7), 760--772 (1998).

\bibitem{Guo2010}
Guo, R., Wong, S.C., Huang, H., Lam, W.H.K.: A microscopic
  pedestrian-simulation model and its application to intersecting flows.
\newblock Phys. A \textbf{389}(3), 515--526 (2010)

\bibitem{Helbing2001}
Helbing, D.: Traffic and related self-driven many-particle systems.
\newblock Rev. Mod. Phys. \textbf{73}, 1067--1141 (2001).

\bibitem{Helbing2000z}
Helbing, D., Farkas, I., Vicsek, T.: Freezing by heating in a driven mesoscopic
  system.
\newblock Phys. Rev. Lett. \textbf{84}, 1240--1243 (2000)

\bibitem{Helbing2006}
Helbing, D., Johansson, A., Mathiesen, J., Jensen, M.H., Hansen, A.: Analytical
  approach to continuous and intermittent bottleneck flows.
\newblock Phys. Rev. Lett. \textbf{97}, 168,001 (2006).

\bibitem{Helbing1995}
Helbing, D., Moln\'ar, P.: Social force model for pedestrian dynamics.
\newblock Phys. Rev. E \textbf{51}, 4282--4286 (1995).

\bibitem{Koester2013}
K\"oster, G., Treml, F., G\"odel, M.: Avoiding numerical pitfalls in social
  force models.
\newblock Phys. Rev. E \textbf{87} (2013)

\bibitem{Maury2007}
Maury, B., Venel, J.: Un mod\`ele de mouvement de foule.
\newblock ESSAIM: Proc. \textbf{18}, 143--152 (2007)

\bibitem{Monneau2013}
Monneau, R., Roussignol, M., Tordeux, A.: Invariance and homogenization of an
  adaptive time gap car-following model.
\newblock Nonlinear Differ. Equ. Appli. \textbf{21}(4), 491--517 (2014)

\bibitem{Nakayama2005}
Nakayama, A., Hasebe, K., Sugiyama, Y.: Instability of pedestrian flow and
  phase structure in a two-dimensional optimal velocity model.
\newblock Phys. Rev. E \textbf{71}, 036,121 (2005)

\bibitem{Ondrej2010}
Ond\^rej, J., Pettr\'e, J., Olivier, A.H., Donikian, S.: A
  synthetic-vision-based steering approach for crowd simulation.
\newblock In: ACM Trans. Graph., vol.~29, pp. 123:1 -- 123:9. ACM, New York,
  NY, USA (2010)

\bibitem{Pelechano2005a}
Pelechano, N., O'Brien, K., Silverman, B., Badler, N.: Crowd simulation
  incorporating agent psychological models, roles and communication.
\newblock In: First International Workshop on Crowd Simulation, vol.~2, pp.
  21--30. Lausanne (2005)

\bibitem{ptv}
{PTV AG}: {PTV Vissim 7.0 -- User Manual}.
\newblock {PTV Group}, Haid-und-Neu-Str. 15, D-76131 Karlsruhe, Germany (2014)

\bibitem{Schadschneider2010a}
Schadschneider, A., Chowdhury, D., Nishinari, K.: Stochastic Transport in
  Complex Systems. From Molecules to Vehicles.
\newblock Elsevier Science Publishing Co Inc. (2010)

\bibitem{Schneider2002}
Schneider, V., K\"onnecke, R.: Simulating evacuation processes with aseri.
\newblock In: Pedestrian and Evacuation Dynamics, pp. 303--314. Springer (2002)

\bibitem{TraffGo2005}
{TraffGo HT GmbH}: Handbuch PedGo 2, PedGo Editor 2 (2005).
\newblock Www.evacuation-simulation.com

\bibitem{Venel2010}
Venel, J.: Integrating strategies in numerical modelling of crowd motion.
\newblock In: Pedestrian and Evacuation Dynamics 2008 (2010)

\bibitem{ZhangJ2012a}
Zhang, J., Klingsch, W., Schadschneider, A., Seyfried, A.: Ordering in
  bidirectional pedestrian flows and its influence on the fundamental diagram.
\newblock J. Stat. Mech. Theor. Exp. \textbf{2012}(02), P02,002 (2012).

\end{thebibliography}

\end{document}